\documentclass[aps,prl,superscriptaddress,showpacs,reprint,floatfix]{revtex4-2}

\usepackage[normalem]{ulem}
\usepackage{graphicx}
\usepackage{color}
\usepackage{xcolor}
\usepackage{amssymb}
\usepackage{amsmath}
\usepackage{comment}
\usepackage{appendix}
\usepackage{lipsum}
\usepackage[makeroom]{cancel}

\usepackage{xcolor}
\definecolor{apsblue}{RGB}{16, 38, 148}
\usepackage[colorlinks=true, allcolors=apsblue]{hyperref}

\newcommand{\ab}{{\alpha\beta}}
\newcommand{\IK}{{\rm IK}}
\newcommand{\bp}{{\bf p}}
\newcommand{\br}{{\bf r}}
\newcommand{\bu}{{\bf u}}
\newcommand{\bF}{{\bf F}}
\newcommand{\bJ}{{\bf J}}

\newcommand{\bnabla}{{\bm{\nabla}}}
\newcommand{\bsigma}{{\bm{\sigma}}}

\newcommand{\Vext}{V_w}
\newcommand{\Vint}{{U}}

\usepackage{amsthm}
\usepackage{bm}
\usepackage[hang,flushmargin]{footmisc}

\usepackage[english]{babel}

\newcommand\rhob{\rho_{\rm b}}

\begin{document}
\author{Jessica Metzger}
\affiliation{Department of Physics, Massachusetts Institute of Technology, Cambridge, MA 02139, USA}
\author{Cory Hargus}
\affiliation{Laboratoire Mati\`ere et Syst\`emes Complexes (MSC), Université Paris Cité  \& CNRS (UMR 7057), 75013 Paris, France}
\affiliation{Entalpic, 5 Parv.\ Alan Turing, 75013 Paris, France}
\author{Julien Tailleur}
\affiliation{Department of Physics, Massachusetts Institute of Technology, Cambridge, MA 02139, USA}
\author{Frédéric van Wijland}
\affiliation{Laboratoire Mati\`ere et Syst\`emes Complexes (MSC), Université Paris Cité  \& CNRS (UMR 7057), 75013 Paris, France}
\affiliation{Yukawa Institute for Theoretical Physics, Kyoto University,
Kitashirakawaoiwake-cho, Sakyo-ku, Kyoto 606-8502, Japan}

\title{Equation of state for the edge flow of chiral colloidal fluids}

\begin{abstract}
We explore the edge flows that emerge at boundaries in nonequilibrium passive and active chiral colloidal fluids.
We show that these complex interface currents obey an equation of state that relates their fluxes to bulk observables.
For confined fluids, the edge flux is given by the average odd stress in the fluid. In phase-separated systems, the flux along the interface is given by the jump of the odd stress across the interface. We then use the equation of state to reveal, and contrast, the microscopic origins of the edge currents in passive and active systems.
\end{abstract}

\maketitle

\noindent
Odd transport phenomena have attracted much interest due to the wealth
of rich behaviors that occur when parity and time-reversal
symmetry are
violated~\cite{lowen_chirality_2016,banerjee_odd_2017,mandal_magnetic_2018,reichhardt_active_2019,epstein_time-reversal_2020,hargus_time_2020,liebchen_chiral_2022,fruchart_odd_2023,deshpande_odd_2024}.
Prominent among them is the emergence of directed boundary flows along
interfaces~\cite{van_zuiden_spatiotemporal_2016,reichhardt_reversibility_2019,ma_dynamical_2022,shen_collective_2023,poggioli_emergent_2023,negi_geometry-induced_2023,caporusso_phase_2024,caprini_self-reverting_2024,adorjani_phase_2024,caprini_bubble_2025,digregorio_phase_2025},
observed in numerous experimental settings, in both
synthetic~\cite{tsai_chiral_2005,tierno_viscoelasticity_2007,yan_rotating_2014,bricard_emergent_2015,workamp_symmetry-reversals_2018,soni_odd_2019,yang_robust_2020,petrichenko_swarming_2020,liu_oscillating_2020,massana-cid_arrested_2021,lopez-castano_chirality_2022,katuri_control_2024,das_lever_2024,caprini_active_2025,zhang_shape-anisotropy_2025}
and
living~\cite{petroff_fast-moving_2015,yamauchi_chirality-driven_2020,beppu_edge_2021,tan_odd_2022,yashunsky_chiral_2022,grober_unconventional_2023,li_robust_2024,grober_hydrodynamics_2025,chao_selective_2026,grober_hydrodynamic_2026}
matter.  These nonequilibrium currents are notoriously
difficult to study and, despite the considerable theoretical work 
dedicated to their characterization~\cite{klymko_statistical_2017,bickmann_analytical_2022,jia_incompressible_2022,kreienkamp_clustering_2022,sese-sansa_microscopic_2022,machado_monteiro_hamiltonian_2023,kalz_field_2024,langford_phase_2025,marconi_emergent_2026,abdoli_dynamical_2026,alsallom_origin_2026,maire_kinetic_2026},
exact laws and generic properties have been scarce, especially in many-body
settings.

In equilibrium, equations of state, exemplified by the ideal gas law,
are powerful tools that yield exact relations between different thermodynamic variables.
Much effort has been devoted to their generalization out of equilibrium, where success has been achieved for static
observables~\cite{bertin_definition_2006,bertin_intensive_2007,palacci_sedimentation_2010,takatori_swim_2014,solon_pressure_2015-1,winkler_virial_2015,ginot_nonequilibrium_2015,speck_ideal_2016,petrelli_effective_2020,hecht_how_2024,caprini_active_2025}. Their application to dynamical observables remains an open challenge.

In this Letter, we tackle this problem for 
odd colloidal fluids comprising either chiral active Brownian particles
(cABPs) or chiral passive Brownian particles (cPBPs) and show their edge flux to obey an equation of state. 
For systems
confined by an external potential, we derive from the microscopics an exact
relation between the net flux along the confining
boundary and the bulk value of the odd stress---the asymmetric contribution to the
generalized stress tensor. Remarkably, this relation implies that, in the macroscopic limit, the edge flux is independent of the details of the confining potential. Analogously, for phase-separated fluids, our theory shows the flux at the
interface between two coexisting phases to be determined by the
difference between their odd stresses.  The fluids we consider consist of $N$ particles in two space dimensions that evolve according to the It\=o-Langevin dynamics:
\begin{align}
    \dot{\br}_i &= \mu \sum_j {\mathbf{F}_{ij}} + \mu \mathbf{f}_i(t) - \mu \bnabla \Vext(\br_i)\;.\label{eq:dynamics-cABP-cPBP}
\end{align}
Here, $\Vext$ is an external potential used to model
confinement, $\mathbf{F}_{ij}$ is the interaction force on
particle $i$ due to particle $j$,
and $\mathbf{f}_i$ is a fluctuating force acting on particle $i$. For
cABPs, $\mathbf{f}_i$ is a propulsion
velocity along an orientation $\bu(\theta_i)$:
\begin{equation}
  \mu \mathbf{f}_i = v_0 \bu(\theta_i)\quad\text{and}\quad \dot{\theta}_i =
  \omega_0 + \sqrt{2 D_r} \xi_i\;,
\end{equation}
with $\xi_i$ a centered unitary Gaussian white noise. We consider conservative and parity-symmetric interactions, $\mathbf{F}_{ij}=-\bnabla_i U^\parallel(r_{ij})$ with $r_{ij}=|\br_{ij}|=|\br_i-\br_j|$, so that chirality arises at the level of the one-body force $\mathbf{f}_i$ when $\omega_0\neq 0$. 
For cPBPs, we use
\begin{align}\label{eq:cPBP-dynamics}
    \mathbf{F}_{ij} &= -\bnabla_i {\Vint}^\parallel_{ij} - \bnabla_i^\perp {\Vint}^\perp_{ij}\quad\text{and}\quad \mathbf{f}_i(t)= \sqrt{2\mu T} \boldsymbol{\eta}_i\;,
\end{align}
where $\bm{\eta}_i$ obeys $\langle \bm{\eta}_i(t) \otimes \bm{\eta}_j(t')\rangle =
2\delta_{ij}\delta(t-t')\mathbb{I}$. Here,
$\Vint^\parallel_{ij}=\Vint^\parallel(r_{ij})$ and
$\Vint^\perp_{ij}=\Vint^\perp(r_{ij})$ are symmetric potentials
which act longitudinally and transversely to the particle separations,
respectively. The transverse forces, which are defined as $-\bnabla_i^\perp \Vint^\perp_{ij}
= -\hat{\mathbf{z}} \times \bnabla \Vint^\perp_{ij}$, have been used to model spinning colloids~\cite{soni_odd_2019,massana-cid_arrested_2021,tan_odd_2022}. They arise from eliminating other degrees of freedom (e.g. internal spins, surrounding fluid) and coarse-graining the corresponding interactions (frictional contacts, hydrodynamic interactions). 
Importantly, transverse forces violate parity
symmetry and induce rotations of interacting pairs. 
Contrary to cABPs, 
the cPBP chirality is  a purely many-body effect arising from $U^\perp$.
At large scales,
cABPs~\cite{liao_clustering_2018,ma_dynamical_2022,langford_phase_2025} and cPBPs ~\cite{hargus_odd_2021,han_fluctuating_2021,vega_reyes_diffusive_2022,caporusso_phase_2024} both diplay odd
diffusion---fluxes transverse to density gradients---and edge
flows. However, they have little in common at the microscopic scale.
It is thus unclear whether their edge flows---one intrinsic, one
emergent---should follow similar laws.

Below, we first show numerically that both systems obey an equation
of state for their edge fluxes. We then detail the case of cABPs,
where we show the edge flux to stem from the asymmetry
of an active stress~\cite{takatori_swim_2014,yang_aggregation_2014,solon_generalized_2018,langford_phase_2025}. We rationalize this asymmetry in terms of the
active impulse---the average momentum transfer from the substrate to
the particle---which has a non-zero component normal to the particle orientation. We then turn to cPBPs, whose
edge flow emerges from the antisymmetric component of the Irving-Kirkwood
stress tensor. We then demonstrate that transverse forces only modify
the steady state of cPBPs due to three-body interactions, which allows us to
accurately predict the edge flux in the dilute limit using an
equilibrium virial expansion. All our numerics are detailed in End
Matter.

\begin{figure}
    \centering
    \includegraphics[width=3.4in]{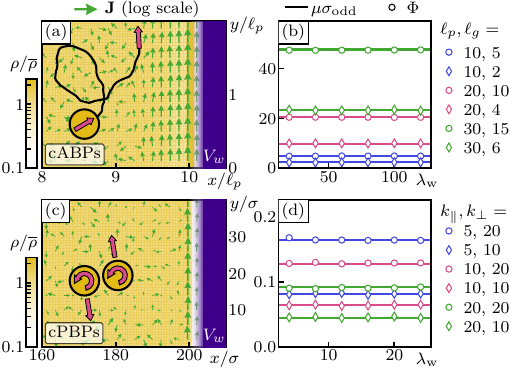}
    \caption{{\bf Edge flux along a confining wall modeled by a repulsive harmonic potential $V_{w}$ of stiffness $\lambda_w$.} {\bf (a)} Density (yellow) and current (arrows) of cABPs confined by an external potential (purple). {\bf (b)} Edge flux (symbols) and odd stress (lines) obey the equation of state~\eqref{eq:EOS-general} for varying persistence lengths $\ell_p=v_0/D_r$, gyroradii $\ell_g=|v_0/\omega_0|$, and wall stiffnesses $\lambda_w$. The odd stress is measured at $\lambda_{\rm w} = 120,24$ for cABPs and cPBPs, respectively. {\bf (c)-(d)} Same as (a)-(b) but for cPBPs, varying the stiffnesses $k_\parallel$, $k_\perp$ of $\Vint^\parallel$, $\Vint^{\perp}$.}
    \label{fig:wall-EOS-demonstration}
\end{figure}

\vspace{0.1in}
\noindent\textbf{\textit{Equation of state for the edge flux.}}  The
average density field of the chiral fluids, defined as
$\rho(\br,t)=\langle \sum_i \delta(\br-\br_i)\rangle$, obeys a
continuity equation $\dot{\rho} = -\partial_\alpha J_\alpha$, where
$\bJ$ is the average particle current. Established
methods~\cite{irving_statistical_1950,kirkwood1949statistical,solon_generalized_2018,langford_phase_2025}
show that, in both active and passive systems, the current can be
written as the divergence of a stress tensor $\bsigma$, i.e.~$J_\alpha
= \mu \partial_\beta \sigma_{\alpha\beta}$. The stress can then be
decomposed between even and odd parts,  $\bsigma = \bsigma_{\rm
  even}+\bsigma_{\rm odd}$ with $\bsigma_{\rm
  even,odd}=(\bsigma\pm \bsigma^{\dagger})/2$. Consider, for simplicity,
a flat interface parallel to the $y$-axis. The flux along this
interface, $\Phi$, is given by the integral of $J_y$ accross the
interface, which satisfies:
\begin{align}
    \Phi &= \int_{x_1}^{x_2} dx J_y = \int_{x_1}^{x_2} dx \mu\big[\partial_x \sigma_{yx} + \partial_y \sigma_{yy}\big]\label{eq:EOS-general1}\\
    &= \int_{x_1}^{x_2} dx \mu\partial_x \sigma_{yx} = \mu\Delta \sigma_{\rm odd}\;.\label{eq:EOS-general}
\end{align}
To get Eq.~\eqref{eq:EOS-general}, we use translational invariance along $\hat{\mathbf{y}}$ and that, in bulk isotropic phases, $\sigma_{\alpha\beta} = \delta_{\alpha\beta} \sigma_{\rm even} + \epsilon_{\alpha\beta} \sigma_{\rm odd}$, where $\epsilon_{\alpha\beta}$ is the Levi-Civita tensor. We denote $\Delta \sigma_{\rm odd}= \sigma_{\rm odd}(x_1)-\sigma_{\rm odd}(x_2)$, where $x_{1,2}$ are abscissas deep in the bulk of each phase. 

\begin{figure}
    \centering
    \includegraphics[width=3.4in]{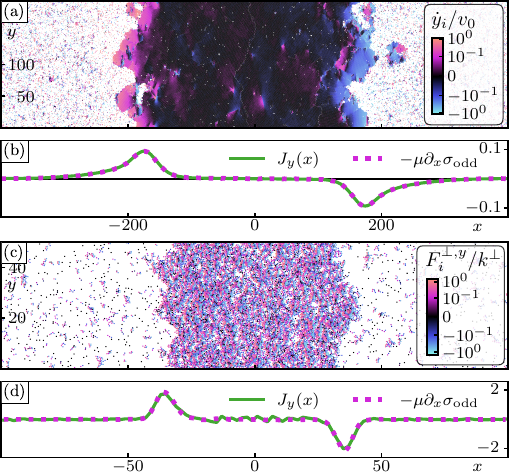}
    \caption{{\bf Boundary flux along emergent interfaces.} {\bf (a)} A motility-induced phase-separated system of cABPs exhibits a steady flow at the phase boundaries. Particles are colored according to the $y$-component of their velocities. {\bf (b)} The edge current stems from the variations of the odd stress, $J_y=-\mu \partial_x \sigma_{\rm odd}$, in agreement with the equation of state~\eqref{eq:EOS-general}. {\bf (c)} Phase-separated cPBPs, colored by the $y$-component of the transverse force they experience. {\bf (d)} The measured current along the slab matches the derivative of the odd stress.
    }
    \label{fig:slab_mips}
\end{figure}

Equation~\eqref{eq:EOS-general} is an equation of state that relates $\Phi$ to $\sigma_{\rm odd}$. It is
agnostic to the nature of the phases separated by the
interface and does not rely on a precise definition of the interface. 
Figure~\ref{fig:wall-EOS-demonstration}  
demonstrates
numerically this equation of state for the flux along a wall, for both
cABPs and cPBPs. In this case, $\bsigma$ vanishes inside the wall and the flux satisfies $\Phi/\mu= \sigma_{\rm odd}$. Since $\sigma_{\rm odd}$ is a bulk property of the system, $\Phi$ is independent of the wall
stiffness, consistent with Figs.\ref{fig:wall-EOS-demonstration}-(b,d). 
In Fig.~\ref{fig:slab_mips}, we consider cABPs undergoing
motility-induced phase separation driven by active and repulsive forces, and cPBPs
undergoing liquid-gas phase separation driven by attractive forces. Both systems are known to exhibit currents along phase
interfaces~\cite{liao_clustering_2018,massana-cid_arrested_2021,ma_dynamical_2022,das_lever_2024,adorjani_phase_2024,caporusso_phase_2024,langford_phase_2025}. Our simulations show the equation of state to
correctly predict not only the total edge flux from the difference between the
odd stresses of the coexisting phases, but also the value of the
transverse current at each point along the interface. Let us now turn to the separate analysis of the odd stress for chiral ABPs and PBPs to compare the microscopic physics at the origin of their edge flows.

\vspace{0.1in}
\noindent\textbf{\textit{Chiral ABPs.}} We consider cABPs interacting via a repulsive potential $U^\parallel$ of typical range $\sigma$. The particle dynamics are then characterized by their dimensionless persistence length $\ell_p=v_0/D_r$ and gyroradius $\ell_g=v_0/\omega_0$. 
Using standard methods~\cite{fily_mechanical_2017,solon_generalized_2018,omar_mechanical_2023}, which have recently been applied to cABPs~\cite{langford_phase_2025}, we find that the particle density $\rho(\br)=\sum_i \langle \delta(\br-\br_i)\rangle$ evolves as $\dot{\rho}=-\partial_\alpha J_\alpha$, with~\footnote{In the presence of particle indices, we move the spatial indices to superscript for clarity.}
\begin{align}\label{eq:currentcABPs}
    \frac{J_\alpha}{\mu} &= \partial_\beta \sigma_\ab^\IK + \frac{v_0}{\mu} \sum_i \langle u^\alpha_i \delta(\br-\br_i)\rangle - \rho \partial_\alpha \Vext\;.
\end{align}
Here $\sigma_\ab^\IK$ is the Irving-Kirkwood stress due to interparticle forces~\cite{irving_statistical_1950}, whose expression for cABPs and cPBPs is derived in End Matter for consistency. The second term on the r.h.s.~is the active force density. Equation~\eqref{eq:currentcABPs} is exact, but not closed. To make quantitative predictions for the flux, one could use approximate closures, as was done for instance in~\cite{solon_generalized_2018,omar_mechanical_2023} for ABPs, in~\cite{poggioli_emergent_2023,langford_phase_2025} for cABPs, or in~\cite{furthauer_active_2013,soni_odd_2019} for cPBPs, at the cost of exactness. 
Instead, here we work directly with Eq.~\eqref{eq:currentcABPs} to analyze the physical ingredients leading to the edge current.

\begin{figure}
    \centering
    \includegraphics[width=3.4in]{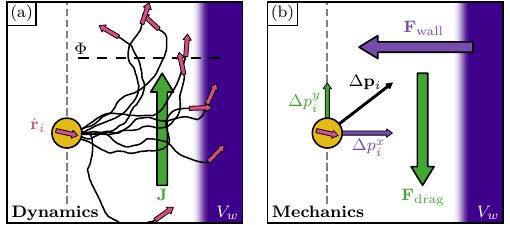}
    \caption{\textbf{Mechanical perspective on the equation of state for cABPs.} {\bf (a)} Sample trajectories of a particle moving towards a wall. The chirality of the dynamics induces a current $\bJ$ along the wall. {\bf (b)} The net momentum gain $\Delta {\bf p}_i$ of the particle is not aligned with its orientation. The component normal to the wall balances the wall force, leading to an active pressure. The component along the wall balances the drag due to the edge flux, leading to the equation of state~\eqref{eq:J-eos-raw}.}
    \label{fig:active-impulse-schematic}
\end{figure}

To proceed, we consider the average momentum an active particle receives in the future, which is given by its `active impulse' $\Delta p_i^\alpha = \int_t^\infty ds \frac{v_0}{\mu}\langle  u^\alpha_i(s)\rangle$~\cite{fily_mechanical_2017}. The active impulse density $\Delta p_\alpha(\br,t)=\sum_i \langle \Delta p_i^\alpha \delta(\br-\br_i)\rangle$ evolves as
\begin{equation}
    \partial_t \Delta p_\alpha = \partial_\beta \sigma^{\rm a}_{\alpha\beta} - \frac{v_0}{\mu} \sum_i \langle u^\alpha_i \delta(\br-\br_i)\rangle\;,~\label{eq:aifield}
\end{equation}
where we have introduced the active stress:
\begin{equation}
 \sigma^{\rm a}_\ab =  -\sum_i \langle \Delta p_i^\alpha \dot{r}_i^\beta \delta(\br-\br_i)\rangle\;.
    \label{eq:sigmaa}
\end{equation}
In the steady state, Eq.~\eqref{eq:aifield} shows the active force density to equal the divergence of the active stress. As a result, all internal forces can be written as the divergence of a stress $\boldsymbol{\sigma}^{\rm tot}=\bsigma^{\rm IK}+\bsigma^{\rm a}$, and the current satisfies $J_\alpha = \mu \partial_\beta \sigma^{\rm tot}_\ab-\mu \rho\partial_\alpha V_w$. 
In the presence of a macroscopic interface along $\hat{\mathbf{y}}$,
projecting this equation along $\hat{\mathbf{y}}$ and integrating along $\hat{\mathbf{x}}$ then leads to Eqs.~\eqref{eq:EOS-general1} and~\eqref{eq:EOS-general}. Since, $\bF_{ij}$ is along $\br_i-\br_j$ for cABPs, $\bsigma^\IK$ is symmetric. The flux of active impulse is thus the sole source of odd stress, $\sigma_{\rm odd}=\sigma^{\rm a}_{xy}$, leading to
\begin{align}
     \Phi_{\rm cABP} &= \mu \Delta \sigma_{\rm odd} = \Big[\mu\sum_i \langle \Delta p_i^y \dot{x}_i \delta(\br-\br_i) \rangle \Big]^{x_1}_{x_2}\;.\label{eq:J-eos-raw}
\end{align}
Let us now give a mechanical interpretation to $\sigma_{\rm odd}$.

\vspace{0.1in}
\noindent
\textbf{\textit{From the  flux of cABPs towards the interface to a momentum flux along it.}} Equation~\eqref{eq:J-eos-raw} relates the particle flux along the interface to the flux of active impulse. 
To compute the latter, we first note that, for fixed $\theta_i(t)$, a particle's expected future orientation satisfies:
\begin{align}
    \langle u_\alpha[\theta_i(t')]\rangle &= e^{-D_r(t'-t)} u_\alpha\big[\theta_i(t) + \omega_0(t'-t)\big]\;.\label{eq:u-rotation}
\end{align}
This allows us to write the active impulse as
\begin{equation}
    \Delta p_i^\alpha(t) = \int_t^{\infty} dt' \frac{v_0}{\mu} \langle u_i^\alpha(t')\rangle = \Omega_{\alpha\gamma} \frac{v_0}{\mu D_r} u_i^\gamma(t)\;,\label{eq:active-impulse-particle}
\end{equation}
where we have introduced the scaled rotation matrix
\begin{equation}
    \Omega_{\alpha\gamma} = \frac{\delta_{\alpha\gamma} - \frac{\omega_0}{D_r} \epsilon_{\alpha\gamma}}{1+\omega_0^2/D_r^2}  = \delta_{\alpha\gamma} \Omega_\parallel - \epsilon_{\alpha\gamma} \Omega_\perp\;.
\end{equation}
The chirality of the dynamics thus rotates the active impulse $\Delta \bp_i(t)$ relative to the orientation $\bu_i(t)$: particles facing an interface later receive a net momentum along it, hence powering $\Phi$. The physical interpretation of Eq.~\eqref{eq:J-eos-raw} is thus a mechanical balance between the net momentum injected along the interface by particles rushing towards it and the drag $\Phi/\mu$ that accompanies the edge flux. This mechanical perspective is illustrated in Fig.~\ref{fig:active-impulse-schematic}.  

\begin{figure}
    \centering
    \includegraphics[width=3.4in]{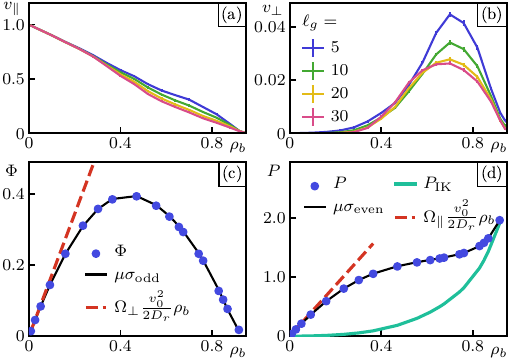}
    \caption{\textbf{Equation of state for cABPs in a confined geometry.} \textbf{(a-b)} Longitudinal and transverse velocities, $v_\parallel$ and $v_\perp$, measured in the bulk ($\ell_p=100$). \textbf{(c)} The flux of cABPs ($\ell_p=10$, $\ell_g=25$) along a wall (blue symbols) agrees quantitatively with the equation of state, Eq.~\eqref{eq:eos-total-flux-interacting} (black line). At low densities, $v_\ab(\rho) \approx v_0 \delta_\ab$ and the ideal equation of state (red dashed line) provides a good approximation of $\Phi$. \textbf{(d)} The mechanical pressure exerted by interacting cABPs on a wall (blue dots) agrees with its prediction by the equation of state in Eq.~\eqref{eq:pressure-eos-interacting} (black line).}
    \label{fig:wall-flux}
\end{figure}

Let us now consider the impact of pairwise forces on the odd stress and edge flow. Using Eq.~\eqref{eq:active-impulse-particle} in Eq.~\eqref{eq:sigmaa} allows us to rewrite the active stress as
\begin{align}
    \sigma^{\rm a}_\ab (\br) &=- \Omega_{\alpha\gamma} \frac{v_0}{2 \mu D_r} \rho(\br) v_{\gamma\beta}(\br)\;,\quad\text{where}\label{eq:v-def}\\
    v_{\alpha\beta}(\br)&=\frac{2}{\rho(\br)} \sum_i \langle u_i^\alpha \dot{r}_i^\beta \delta(\br-\br_i)\rangle\;
\end{align}
is a tensorial generalization of the effective speed introduced to study active particles interacting via pairwise forces~\cite{fily_athermal_2012,stenhammar_phase_2014,berthier_glassy_2019,arnoulx_de_pirey_active_2019}.
In an isotropic bulk with density $\rhob$, it acquires an intuitive form $v_{\alpha\beta}=v_\parallel \delta_{\alpha\beta}-v_\perp \epsilon_{\alpha\beta}$, where
\begin{align}
   v_\parallel(\br_b) &= v_0 + \frac{\mu}{\rhob} \sum_i \langle \mathbf{F}_i \cdot \bu_i \delta(\br_b-\br_i)\rangle\label{eq:vpar}\\
     v_\perp(\br_b) &= \frac{\mu}{\rhob} \sum_i \langle \mathbf{F}_i \cdot \bu_i^\perp \delta(\br_b-\br_i)\rangle\;,\label{eq:vperp}
\end{align}
with $\mathbf{F}_i=\sum_j \mathbf{F}_{ij}$, $\br_b$  a location in the bulk, and $u_{i,\alpha}^\perp = \epsilon_{\alpha\beta} u_{i,\beta}$. 
The effective velocity $v_\parallel$ measures the average particle velocity along their orientations~\cite{fily_athermal_2012,stenhammar_phase_2014,berthier_glassy_2019,arnoulx_de_pirey_active_2019}. Its decrease at high density drives motility-induced phase separations in achiral~\cite{tailleur_statistical_2008} and chiral~\cite{sese-sansa_microscopic_2022,langford_phase_2025} active matter. 
The effective velocity $v_\perp$ instead measures the average particle velocities \textit{normal} to their orientations. Its nonvanishing in an isotropic bulk is a signature of chirality. Together, $v_\perp$ and $v_\parallel$ allow rewriting the active stress as
\begin{align}
    \sigma^{\rm a}_\ab= \frac{v_0 {\rhob}}{2\mu D_r} \Big[\epsilon_\ab \big(\Omega_\perp v_\parallel + \Omega_\parallel v_\perp\big)-\delta_\ab \big(\Omega_\parallel v_\parallel - \Omega_\perp v_\perp\big)\Big]\;.\label{eq:active-impulse-flux-Omega-v}
\end{align}

Equation~\eqref{eq:active-impulse-flux-Omega-v} naturally splits the active stress between odd and even parts, which allows us to identify their microscopic origins. Combining Eqs.~\eqref{eq:J-eos-raw} and~\eqref{eq:active-impulse-flux-Omega-v}, the odd stress shows the edge flux to be given by
\begin{align}
    \Phi_{\rm cABP} &= \frac{v_0 {\rhob}}{2 D_r} \big[\Omega_\perp v_\parallel + \Omega_\parallel v_\perp\big]^{x_2}_{x_1}\;.\label{eq:eos-total-flux-interacting}
\end{align}
Two effects thus power the edge current: first, the rotation in the $\perp$ direction of $v_\parallel$ due to $\omega_0$; and second, the transport in the $\parallel$ direction of $v_\perp$, which stems from non-vanishing average interparticle forces normal to the particle orientations. 
We numerically measure $v_\parallel$ and $v_\perp$  in Fig.~\ref{fig:wall-flux}(a)-(b) for varying $\omega_0$ and $\rhob$. While $v_\perp$ is weak compared to $v_\parallel$, it increases with $\omega_0$ and is peaked at a finite density. We then use Eq.~\eqref{eq:eos-total-flux-interacting} to generate predictions for the edge flux along a confining wall, shown as the black line in Fig.~\ref{fig:wall-flux}(c). Simultaneously, we directly measure the flux along the wall, and report quantitative agreement with the equation-of-state predictions, hence demonstrating our mechanical interpretation of the edge flux and its bulk-edge correspondence.

Equation~\eqref{eq:active-impulse-flux-Omega-v}  also yields insight into the pressure of cABPs~\cite{caporusso_phase_2024,langford_phase_2025,caprini_active_2025}. It predicts that the force density exerted on the confining wall can be measured in the bulk of the system as
\begin{align}
    P &= \frac{v_0 \rho_b}{2 D_r} \big( \Omega_\parallel v_\parallel - \Omega_\perp v_\perp\big) - \sigma_{xx}^\IK \;.\label{eq:pressure-eos-interacting}
\end{align}
This time, the pressure arises from the contribution of $v_\parallel$ which is not rotated away from the particle orientation, and, more surprisingly, from the contribution of $v_\perp$ which is rotated back  along the particle orientation by $\omega_0$. Figure~\ref{fig:wall-flux}(d) shows a quantitative agreement between this bulk prediction  and the measurement of pressure as a force density on the confining wall. We now turn to cPBPs, where chirality is an emergent property due to transverse interactions, rather than a direct consequence of a chiral one-body self-propulsion.

\begin{figure}
    \centering
    \includegraphics[width=3.4in]{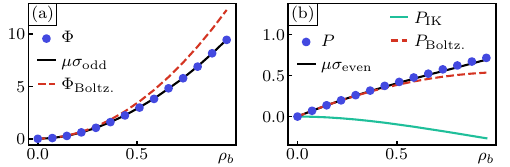}
    \caption{\textbf{The edge flux induced by transverse forces in cPBPs obeys an equation of state}. {\bf (a)} The edge flux induced by a confining wall in interacting cPBPs. The directly measured flux (blue symbols) agrees with the equation of state prediction~\eqref{eq:cpbp-phi-eos} (black line). At low densities, the dilute Boltzmann approximation (red line) also agrees. {\bf (b)} Same as (a), but for the pressure exerted on the wall. }
    \label{fig:cPBP}
\end{figure}

\vspace{0.1in}
\noindent\textbf{\textit{Chiral PBPs.}} 
Stochastic calculus~\cite{dean_langevin_1996} shows  Eq.~\eqref{eq:cPBP-dynamics} to imply an evolution of the density field of the form
\begin{align}
    \dot\rho =&-\bnabla \cdot \big[\mu \bnabla\cdot\bsigma^\IK - \mu T\bnabla \rho -\mu \rho\bnabla\Vext\big]\;,\label{eq:rhodot-cPBP}
\end{align}
where established methods~\cite{irving_statistical_1950,klymko_statistical_2017}, detailed in End Matter, show the Irving-Kirkwood stress tensor to read~\footnote{Note that Eq.~\eqref{eq:sigmaIK-cPBP} is not exact without the integral on $\lambda$, except in a homogeneous bulk.}
\begin{align}
    \bsigma^\IK(\br) = -\sum_{i<j} \bigg\langle \mathbf{F}_{ij} \otimes \br_{ij} \int_0^1 d\lambda \delta(\br-\br_i + \lambda \br_{ij})\bigg\rangle\;.\label{eq:sigmaIK-cPBP}
\end{align}
Notably, all pairwise forces---including transverse ones---can be written as the divergence of  $\bsigma^\IK$. The equation of state~\eqref{eq:EOS-general} then tells us that the edge flux is entirely determined by its odd component. In a homogeneous phase, Eq.~\eqref{eq:sigmaIK-cPBP} yields
\begin{align}
    \Phi_{\rm cPBP} &= \frac{\mu }{2 } \Big[\sum_{i} \langle \delta(\br-\br_i) x_{ij} \partial_{x_i} \Vint^\perp_{ij} \rangle \Big]^{x_2}_{x_1}\;.\label{eq:cpbp-phi-eos}
\end{align}
Equation~\eqref{eq:cpbp-phi-eos} confirms that the edge flux originates in the transverse forces. To get more insight, we note that, at low densities, transverse forces barely modify the $N$-body probability density $\mathcal P(\br_1,\ldots,\br_N,t)$. Consider indeed the Fokker-Planck equation $\dot {\mathcal{P}}=-H\mathcal {P}$, with:
\begin{equation}
    H = - \sum_i \bnabla_i \cdot \bigg[T \bnabla_i + \sum_j \big(\bnabla_i \Vint^\parallel_{ij} + \bnabla_i^\perp \Vint^\perp_{ij} \big)\bigg]\;.
\end{equation}
The Boltzmann distribution $\mathcal P_{\rm eq} \propto e^{-\sum_{i<j} {\Vint^\parallel_{ij}}/T}$ satisfies
\begin{equation*}
    \frac{H \mathcal{P}_{\rm eq}}{\mathcal P_{\rm eq}} = \sum_{i,j,k} \frac{\bnabla_i \Vint^\parallel_{ik} \cdot \bnabla_i^\perp \Vint^\perp_{ij}}{T} =- \sum_{i,j,k} \frac{\hat{\mathbf{r}}_{ik} \times \hat{\mathbf{r}}_{ij} (\Vint^\parallel_{ik})' (\Vint^\perp_{ij})'}{T}\;.\label{eq:cPBP-Boltzmann}
\end{equation*}
The summand is nonzero only for $U^\parallel\neq U^\perp$, and for $j\neq k$, i.e.~during 3-body collisions. It can thus be neglected in dilute mixtures, thus allowing for a quantitative prediction of $\Phi_{\rm cPBP}$.

In Fig.~\ref{fig:cPBP}(a-b), we report numerical simulations of cPBPs with varying densities confined by a vertical wall potential. The numerical measurement of $\sigma_{\rm odd}$ using Eq.~\eqref{eq:cpbp-phi-eos} agrees perfectly with the direct measurement of the particle flux along the wall, consistently with our derived exact equation of state. Using a virial approximation of the Boltzmann pair correlation function, $P_2(\br,\br') \approx \rho_b^2 e^{-\Vint^\parallel(\br-\br')/T}$, we then calculate a theoretical estimates $\Phi_{\rm Boltz.}$, as detailed in End Matter, which matches our numerics at low densities. Finally, our approach can also be used to predict $\sigma_{\rm even}$ at low density, and thus the pressure exerted by cPBPs on the wall, as shown in Fig.~\ref{fig:cPBP}(b).

\vspace{0.1in}
\noindent\textbf{\textit{Conclusion.}} 
Linear response allows relating fluxes to infinitesimal thermodynamic driving, in~\cite{kubo_fluctuation-dissipation_1966} and out~\cite{agarwal_fluctuation-dissipation_1972,baiesi_update_2013} of equilibrium. 
Here, we uncover an exact relation between the edge flux of chiral colloidal fluids and their bulk stresses valid for arbitrary drive strength. It takes the form of an equation of state, which allows us to decipher the microscopic physics powering edge flows in passive and active chiral Brownian particles. 

Our results can now be extended to many different settings, such as underdamped cPBPs. These are  particularly interesting due to their exotic phenomenology ranging from bubble formation~\cite{shen_collective_2023,caprini_bubble_2025} and ``chiral liquid" phases~\cite{bililign_motile_2022,caporusso_phase_2024} to arrested coarsening~\cite{massana-cid_arrested_2021}. 
Another important direction where currents play a prominent role is the study of interfacial, wetting, and capillary phenomena~\cite{zakine_surface_2020,wysocki_capillary_2020,fins_carreira_how_2024,zhao_wetting_2026}. How edge currents modify their physics is a largely open question, where our equation-of-state formalism could be put to work.

\vspace{0.1in}
\noindent\textbf{\textit{Acknowledgements.}}

We thank F. Ghimenti for useful discussions. FvW and JT thank the Agence Nationale de la Recherche grant THEMA No. 20-CE30-0031-01 grant. JT and JM thank the laboratoire Matière and Systèmes Complexes for hospitality and MISTI GSF for funding. JM thanks ICAM-I2CAM Institute for Complex Adaptive Matter for funding.

\vspace{0.1in}
\noindent\textbf{\textit{End matter.}} Below, we first provide the details of our numerical simulations (integration scheme, interaction forces, boundary conditions, and simulation parameters) for all figures before detailing the derivation of $\bsigma^{\rm IK}$ and of the virial fluxes and pressure shown in Fig.~\ref{fig:cPBP}.

\vspace{0.1in}
\noindent\textit{Integration scheme.} 
All simulations use an Euler-Maruyama integration scheme. For Figs.~\ref{fig:slab_mips}(a-b) and~\ref{fig:wall-flux}, we used LAMMPS's Euler integrator with adaptive time-stepping.

\vspace{0.1in}
\noindent\textit{Interaction forces.} 
For the cABP simulations shown in Fig.~\ref{fig:wall-EOS-demonstration}, we use a repulsive harmonic potential $\Vint^\parallel(r)=\frac{k\sigma}{2} \big(1-\frac{r}{\sigma}\big)^2 \Theta(\sigma-r)$, where $\Theta(x)$ is the Heaviside step function. For the cABP simulations shown in Figs.~\ref{fig:slab_mips} and~\ref{fig:wall-flux}, we use a purely repulsive WCA potential $\Vint^\parallel(r)=\big\{4\epsilon \big[(\sigma/r)^{12}-(\sigma/r)^6\big] + \epsilon\big\}\Theta(2^{1/6} \sigma-r)$.

For the cPBP simulations shown in Fig.~\ref{fig:cPBP}, we use a potential $U^\parallel$ that is attractive, with an equilibrium separation $r_0$, and has a soft repulsive core:
\begin{align}
    U^\parallel(r) &= k_\parallel \frac{\frac{-2r^3}{\sigma^3} + \frac{3 r^2}{\sigma^2} \big(1+\frac{r_0}{\sigma}\big) - \frac{6r}{\sigma} + \frac{3 r_0}{\sigma} - 1}{\frac{3r_0}{\sigma}-1} \Theta(\sigma-r)\;.\label{eq:cubic-potential}
\end{align}
For Fig.~\ref{fig:wall-EOS-demonstration}(c-d), we use the repulsive harmonic potential $U^\parallel(r) = k_\parallel \frac{\sigma}{2} \big(1-\frac{r}{\sigma}\big)^2 \Theta(\sigma-r)$.

For transverse interactions between cPBPs, we always use the repulsive harmonic potential $U^\perp(r) = k_\perp \frac{\sigma}{2} \big(1-\frac{r}{\sigma}\big)^2 \Theta(\sigma-r)$. We use the same $\sigma$ for $U^\parallel$ and $U^\perp$.

\vspace{0.1in}
\noindent\textit{Boundary conditions and external potential.}
The simulations in Figs.~\ref{fig:slab_mips} and~\ref{fig:wall-flux}(a-b) use periodic boundary conditions in the $x$ and $y$ dimensions. The simulations shown in Figs.~\ref{fig:wall-EOS-demonstration},~\ref{fig:wall-flux}(c-d), and~\ref{fig:cPBP} use periodic conditions in the $y$ dimension and closed boundary conditions in the $x$ dimension. In Figs.~\ref{fig:wall-EOS-demonstration} and~\ref{fig:cPBP}, the closed boundary conditions are enforced by a wall modeled using the harmonic potential
\begin{align}
    \Vext(x) &= \begin{cases}
        \frac{\lambda_w}{2} x^2\;,\; &x\leq 0\\
        \frac{\lambda_w}{2} \big(x-L_x\big)^2\;,\; &x\geq L_x\\
        0\;,\;&\text{otherwise}
    \end{cases}\;.
\end{align}
In Fig.~\ref{fig:wall-flux}(c-d), the wall is modeled using a repulsive WCA potential.

\vspace{0.1in}
\noindent\textit{Simulation parameters.}
For all simulations, we use an interaction radius $\sigma=1$ and a mobility $\mu=1$.

 \vspace{0.1in}
\noindent\textit{Figure~\ref{fig:wall-EOS-demonstration}.} 
Fig.~\ref{fig:wall-EOS-demonstration}(a-b) consists of simulations of $N=2400$ cABPs in a system of size $L_x=200\sigma$, $L_y=40\sigma$, interacting via a repulsive harmonic interaction potential with strength $k=60$. In panel (a), we take $\ell_p=20$, $\ell_g=10$, and $\lambda_w=20$. Fig.~\ref{fig:wall-EOS-demonstration}(c-d) shows simulations of $N=2400$ cPBPs in a system of size $L_x=200\sigma$ and $L_y=40\sigma$ interacting via a repulsive harmonic longitudinal interaction potential and harmonic transverse interaction potential. In panel (c), we take $k_\parallel=5$, $k_\perp=20$, and $\lambda_w=4$. In all panels, we use closed boundary conditions in the $x$ direction with a quadratic wall of varying strength $\lambda_w$. 
The agreement found between the current and the odd stress for cABPs in Fig.~\ref{fig:wall-EOS-demonstration} was obtained by measuring $\sigma_{xy}$ from Eq.~\eqref{eq:sigmaa}. 

 \vspace{0.1in}
\noindent\textit{Figure~\ref{fig:slab_mips}.} 
In Fig.~\ref{fig:slab_mips}(a-b), we simulate $N=71,442$ cABPs interacting via a WCA potential of strength $\epsilon=1$, in a system of size $L_x=800\sigma$ and $L_y=200\sigma$. We take $\ell_p = 200$ and $\ell_g = 100$. In Fig.~\ref{fig:slab_mips}(c-d), we simulate $N=13,500$ cPBPs in a system of size $L_x=200\sigma$ and $L_y=50\sigma$ interacting via a cubic longitudinal potential [Eq.~\eqref{eq:cubic-potential}] of strength $k_\parallel=50$ and equilibrium separation $r_0=0.667$, and a harmonic transverse potential of strength $k_\perp=12$. 
The agreement found between the current and the odd stress for cABPs in Fig.~\ref{fig:slab_mips} was obtained by measuring $\sigma_{xy}$ from Eq.~\eqref{eq:sigmaa}. 

 \vspace{0.1in}
\noindent\textit{Figure~\ref{fig:wall-flux}.} 
Fig.~\ref{fig:wall-flux} shows the results of simulations of cABPs interacting via a WCA potential of strength $\epsilon=$1, with number densities $\rho_b \in [0.008,0.92]$ (x-axes). In Fig.~\ref{fig:wall-flux}(a-b), we use a system of size $L_x=L_y=200\sigma$, and take $\ell_p=100$ while varying $\ell_g$. In Fig.~\ref{fig:wall-flux}(c-d), we use a system of size $L_x=100\sigma$ and $L_y=40\sigma$, with $\ell_p=10$ and $\ell_g=25$. We place a wall on the $x$ boundaries which interacts with particles through a WCA potential.

 \vspace{0.1in}
\noindent\textit{Figure~\ref{fig:cPBP}.} 
In Fig.~\ref{fig:cPBP}, we simulate $N\in \{500,1000,\ldots,4500\}$ cPBPs in a system of size $L_x=64\sigma$ and $L_y=32\sigma$ confined in the $x$ direction by a quadratic wall of strength $\lambda_w=20$. The cPBPs have interaction strengths $k_\parallel=40$ and $k_\perp=80$, and equilibrium separation $r_0=0.7$.

\vspace{0.1in}
\noindent\textit{Derivation of $\bsigma^{\rm IK}$.}
Let us now prove that all the forces, including the transverse ones, can be written as the divergence of a stress tensor $\bsigma^\IK$. The contribution of the interaction forces to the particle current at $\br$ is given by
\begin{align}
    \sum_{i,j} \bF_{ij} \delta(\br-\br_i) &= \frac12 \sum_{i,j} \big(\bF_{ij}-\bF_{ji}\big) \delta(\br-\br_i),\label{eq:sigIK-cpbp-deriv1}
\end{align}
where we have used the fact that all forces satisfy action-reaction; i.e.
\begin{align}
    \bF_{ij} &= -\bnabla_i \Vint^\parallel(r_{ij}) - \hat{\mathbf{z}} \times \bnabla_i \Vint^\perp(r_{ij}) \\
    &= \bnabla_j \Vint^\parallel(r_{ij}) + \hat{\mathbf{z}} \times \bnabla_j \Vint^\perp(r_{ij}) = -\bF_{ji}\;.\label{eq:sigIK-cpbp-deriv2}
\end{align}
We then find
\begin{align}
    \sum_{i,j} &\bF_{ij} \delta(\br-\br_i) = \frac12 \sum_{i,j} \bF_{ij} \big[\delta(\br-\br_i)-\delta(\br-\br_i)\big]\\
    &= -\frac12 \bnabla \cdot \sum_{i,j} \bF_{ij} \otimes \br_{ij} \int_0^1 d\lambda \delta(\br-\br_i+\lambda\br_{ij})\\
    &= \bnabla \cdot \bsigma^\IK\;,\label{eq:sigIK-cpbp-deriv3}
\end{align}
where in the first equality we have reindexed the right-hand side of Eq.~\eqref{eq:sigIK-cpbp-deriv1}, and in the second we have used the identity~\cite{kirkwood1949statistical}
\begin{align}
    \delta(\br-\br_i)&-\delta(\br-\br_j) \\
    &= -\bnabla \cdot \bigg[\br_{ij} \int_0^1 d\lambda \delta(\br-(1-\lambda)\br_i-\lambda\br_j)\bigg]\;.\nonumber
\end{align}
From Eq.~\eqref{eq:sigIK-cpbp-deriv3}, we thus arrive at Eq.~\eqref{eq:sigmaIK-cPBP}.

\vspace{0.1in}
\noindent\textit{Theoretical curves in Fig.~\ref{fig:cPBP}.}
In Fig.~\ref{fig:cPBP}, the theoretical flux $\Phi_{\rm Boltz.}$ and pressure $P_{\rm Boltz.}$ (red dashed lines) are constructed as follows. We use the virial approximation to write the pair correlation function as $P_2(\br,\br') \approx \rho_b^2 e^{-\Vint^\parallel(\br-\br')/T}$~\cite{hansen_theory_2013}, from which we calculate the Irving-Kirkwood stress using a variant of Eq.~\eqref{eq:sigmaIK-cPBP}:
\begin{align}
    \bsigma_\IK &= \int d\br' \mathbf{F}(\br') \otimes \br' \int_0^1 d\lambda \frac{P_2\big(\br+(1-\lambda)\br',\br-\lambda \br'\big)}{2}\nonumber\\
    &\approx -\frac{\rho_b^2}{2} \int d\br' \big[\bnabla U^\parallel + \bnabla^\perp U^\perp\big] \otimes \br' e^{-\Vint(r')^\parallel/T} \;,
\end{align}
where we have used $\mathbf{F}=-\bnabla U^\parallel - \bnabla^\perp U^\perp$. This is then numerically integrated to find the result.

\bibliography{refs}

\end{document}